# Slip-link simulations of long-fiber networks under uniaxial compression

Yuichi Masubuchi*

*Department of Materials Physics, Nagoya University, Nagoya 464-8603, Japan*

*Corresponding author: e-mail: mas@mp.pse.nagoya-u.jp

**Abstract:** A coarse-grained molecular simulation approach originally developed for entangled polymeric liquids is extended to model the mechanical behavior of long-fiber networks. The model, based on the slip-link picture of chain entanglements, resolves the force balance at contact points and accounts for fiber slippage under these topological constraints. Two key governing equations describe the time evolution of contact-point positions and the local fiber fraction between adjacent contact points. A yield-force criterion determines whether contact points are displaced or remain pinned, as well as whether fiber slippage occurs at contact points. Uniaxial compression simulations corresponding to press molding of fiber-reinforced thermoplastics were performed for networks with varying fiber lengths and compression rates. The results were qualitatively consistent with experimental observations of long-fiber thermoplastics. The model captures physics inaccessible to the classical van Wyk theory of fiber network compression, which is quasi-static and insensitive to fiber length. This work demonstrates that the slip-link framework, already validated for polymer melts, provides a promising mesoscale simulation tool for understanding and predicting the processing behavior of non-thermal fiber networks.

## 1. Introduction

Long-fiber-reinforced thermoplastics (LFTs) and glass-mat thermoplastics (GMTs) are increasingly used in automotive and structural applications for their excellent strength-to-weight ratio, impact resistance, and recyclability[1], [2]. In compression molding of these materials, a charge of intertwined long fibers embedded in a thermoplastic matrix is pressed into a mold cavity, where the material must flow to fill the desired geometry [3], [4], [5]. A fundamental challenge in this process is controlling the flow of fibers without excessive breakage; retaining fiber length is critical for the mechanical performance of the final part, yet longer fibers resist flow more strongly due to inter-fiber contacts [6].

The rheological behavior of fiber-reinforced materials during compression molding has been studied extensively through squeeze-flow experiments. Kotsikos et al. [7] characterized the squeeze flow of GMT and demonstrated strongly anisotropic, rate-dependent behavior. Rienesl et al. [8] developed squeeze-flow rheometry methods for carbon-fiber sheet molding compounds (CF-SMC) and calibrated power-law viscosity models. More recently, Schreyer et al. [9] examined the anisotropic flow behavior of long carbon-fiber-reinforced polyamide 6 using isothermal squeeze-flow tests at various temperatures and compression velocities, reporting a pronounced shear-rate dependence. Squeeze-flow studies on E-glass/polypropylene LFTs by Dweib and ÓBrádaigh [10] systematically investigated the effects of fiber length, weight fraction, and temperature on the rheological response. A recurrent observation across these studies is that longer fibers and higher fiber concentrations increase the resistance to flow, and that the stress response is strongly rate-dependent.

An important phenomenon observed in GMT compression molding is fiber–matrix separation: the polymer matrix flows more readily than the fiber network, leading to spatial inhomogeneity in fiber content [11]. Huang et al. [12] documented this effect experimentally and showed that the power-law index changes progressively during compression as the internal fiber microstructure evolves. This observation highlights the central role of inter-fiber entanglements in governing the macroscopic flow behavior.

On the theoretical side, the compression mechanics of fiber networks has a long history dating back to van Wyk [13], who modeled the compression of wool as bending of fiber segments supported at inter-fiber contact points. The van Wyk theory predicts that compressive stress scales as the third power of the fiber volume fraction, $\sigma \propto \varphi^3$, and depends on the fiber elastic modulus and diameter, but is independent of fiber length and insensitive to compression rate. Toll [14] extended this framework to account for preferential fiber alignment, predicting a higher power-law exponent. More recently, Picu and Negi [15] revisited the stress–density relation using numerical simulations and identified regimes beyond the classical van Wyk prediction. Toll and Manson [16] provided an elastic compression theory for planar fiber networks. Despite these advances, the quasi-static, purely elastic nature of these theories means they cannot capture rate-dependent or fiber-length-dependent flow phenomena that are central to compression molding.

Numerical simulation of fiber networks has advanced considerably. Abd El-Rahman and Tucker [17] performed direct finite-element simulations of random long-fiber network compression using 3D beam elements, resolving individual fiber–fiber contacts for systems of 5000 fibers. Their results agreed with van Wyk theory at low-to-moderate densities but revealed deviations at higher densities and demonstrated the evolution of fiber orientation during compression. Discrete element method (DEM) simulations by Guo et al. [18] investigated the effects of fiber shape (curvature) on compression, tension, and shear behavior of fiber assemblies. Negi and Picu [19] studied non-crosslinked fiber networks with adhesion and friction. While these approaches capture rich micro-mechanical detail, they are computationally demanding and difficult to extend to incorporate dynamic, rate-dependent entanglement phenomena.

Meanwhile, in the polymer physics community, coarse-grained molecular models for the multi-body dynamics of threadlike molecules have been developed [20], [21], [22], [23]. Although a crucial difference is the absence of thermal agitation in macroscopic fiber networks, the topological features of entwined threads share fundamental similarities; thus, frameworks developed for polymers are likely applicable to macroscopic fiber systems with appropriate modifications. For instance, the multi-chain slip-link network model [24] represents entangled polymers as a network of nodes, strands, and dangling ends. This construction can be adapted for macroscopic fibers by modifying the equation of motion.

This Letter proposes an extension of the multi-chain slip-link model to simulate the uniaxial compression of long-fiber networks. The model naturally captures both fiber-length dependence and rate dependence of the compressive stress—physics that are inaccessible to van Wyk-type theories. The results are qualitatively compared with experimental observations from LFT and GMT compression molding.

## 2. Model

The simulation model is based on the primitive chain network (PCN) framework [24] adapted here for fiber networks. Each fiber is represented as a sequence of contact points that topologically constrain the fiber motion. The contact points are modeled as slip-links that maintain contact between fibers until the fibers slide apart. The fibers are assumed to be well dispersed (opened) and unbundled. Figure 1 schematically shows the model. The state variables are the position vectors $\{\mathbf{R}\}$ of the slip-links, the local fiber fraction $\{n\}$ between adjacent slip-links along each fiber, and the number of slip-links on the fibers $\{Z\}$.

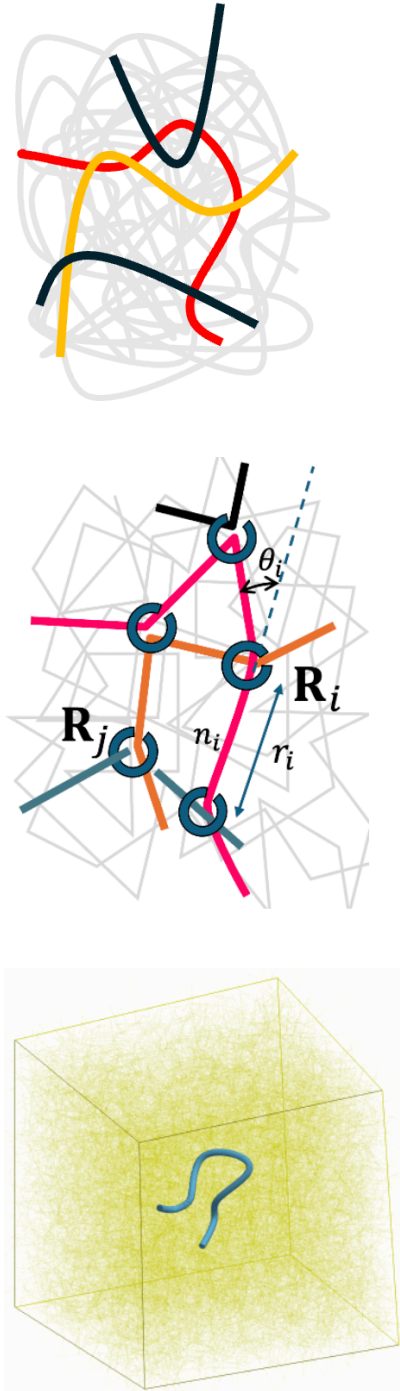

**Figure 1** A schematic representation of a long-fiber matrix (top), the slip-link model (mid), and a typical snapshot of the simulation (bottom). In the bottom panel, one of the fibers is shown as a bold blue spline curve connecting 10 network nodes, and the others as thin yellow lines.

The time evolution of the slip-link position $\mathbf{R}_i$ is governed by a force balance:

$$\mathbf{0} = -\zeta\left(\frac{d\mathbf{R}_i}{dt} - \boldsymbol{\kappa} \cdot \mathbf{R}_i\right) + \mathbf{F}_i \tag{1}$$

where $\zeta$ is a friction coefficient for spatial displacement, $\boldsymbol{\kappa}$ is the strain rate tensor, and $\mathbf{F}_i$ is the net force arising from bonding and bending interactions written below.

$$U = U_e + U_b \tag{2}$$

$$U_e = \frac{1}{2}k_e \sum_i \left(\frac{r_i}{n_i} - \frac{r_0}{n_0}\right)^2 \tag{3}$$

$$U_b = \frac{1}{2}k_b \sum_i (\cos\theta_i - 1)^2 \tag{4}$$

Here, $k_e$ is the spring constant, $r_i$ is the length of the fiber segment (network strand), and $n_i$ is the fiber fraction that describes fiber sliding as explained later. $r_0$ and $n_0$ are parameters to control the equilibrium network mesh size. $k_b$ is the intensity parameter of the bending potential, and $\theta_i$ is the bending angle between adjacent segments.

In addition to the slip-link position, fiber sliding is described by the change of the fiber fraction $n_i$ on each network strand.

$$0 = -\xi_s \frac{dn_i}{dt} + f_i \tag{5}$$

where $\xi_s$ is a friction coefficient for fiber sliding, and $f_i$ is the driving force for contour redistribution along the fiber.

Numerically solving equations 1 and 5 corresponds to the energy-minimization scheme called the Broyden–Fletcher–Goldfarb–Shanno method [25]. The network structure obtained by minimizing the total potential $U$ is a mesh of straight rods that resists deformation with zero stress owing to fiber sliding; such a state is insufficient as a model for long-fiber composites. To prevent this network collapse, a yield-force criterion is applied. Namely, for the slip-link position $\mathbf{R}_i$, the test force is obtained by $\mathbf{F}_t = -\nabla U$, and $\mathbf{F}_i = \mathbf{F}_t$ when $|\mathbf{F}_t| > F_y$, where $F_y$ is a threshold parameter that mimics the static friction at fiber–fiber contact points. Otherwise, $\mathbf{F}_i = \mathbf{0}$, and the slip-link moves affinely according to $\boldsymbol{\kappa}$. A similar rule applies to the fiber fraction $n_i$. The scalar trial force $f_t$ is the balance of elastic tension between two adjacent segments, and $f_i = 0$ if $|f_t| < F_y$, and $f_i = f_t$ otherwise.

In addition to the fiber motion mentioned above, network rearrangement at the dangling ends is considered. At each dangling end, $n_e$ is monitored if the number is within a preset window; $1/2 < n_e/n_0 < 3/2$. If $n_e$ falls below the minimum threshold, as the fiber slides off the contacting fiber, the connecting slip-link is removed, and the contacting fiber is released. Conversely, if $n_e$ exceeds the maximum value, the fiber protrudes from the contact point and contacts the other fiber, creating a new slip-link.

The simulations were performed in a three-dimensional box with periodic boundary conditions. Fibers were initially placed in random configurations at a fixed volume fraction, and the system was relaxed until $\{\mathbf{R}\}$, $\{n\}$, and $\{Z\}$ became stable. Figure 1 bottom panel shows an example of such a relaxed network. Then, uniaxial compression was imposed by shrinking the box dimension in the compression direction at a constant rate, while expanding the two transverse dimensions to maintain constant volume (corresponding to biaxial extension in the transverse plane). This geometry corresponds to the press-molding process used in LFT and GMT manufacturing [2], [4].

Units of length, energy, and time are chosen as $r_0$, $k_e r_0^2/n_0$, and $\zeta n_0/k_e$. The other quantities are normalized according to these units. The friction ratio was $\zeta_s/\zeta = 1/5$, the bending parameter was $k_b/k_e = 5$, and the yield force was chosen at $F_y = 0.1$. The simulation box dimension was $16^3$, and the initial number density of strands (before relaxation) was 10. The fiber length $n_t$ and the compression strain rate $\dot{\varepsilon}_b$ (Hencky biaxial strain rate) were varied, and their effects on the stress response are shown below.

## 3. Results

Figure 2 shows the compressive stress as a function of the deformation ratio $d/d_0$ (where $d_0$ is the initial thickness) for networks with $n_t/n_0$ =3, 5, 10, and 20. For short fibers ($n_t/n_0$ =3), the stress $\sigma$ remains low throughout compression, indicating that the network flows readily after yielding. As $n_t/n_0$ increases, the stress rises significantly. The stress difference between $n_t/n_0$ =3 and 20 spans approximately two orders of magnitude. This strong dependence on $n_t/n_0$ demonstrates that inter-fiber contacts are the dominant factor controlling compressive resistance.

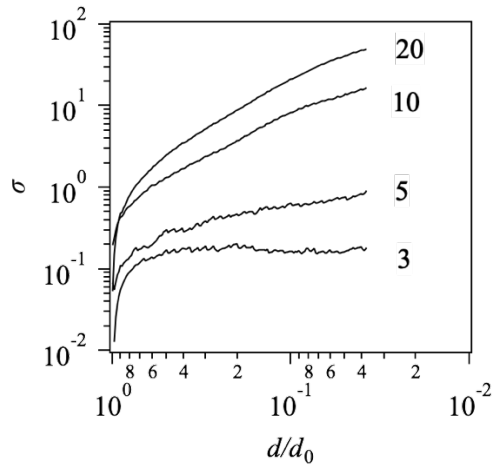

**Figure 2** Stress response under compression for the fiber networks with various fiber lengths $n_t/n_0$ at the compression rate $\dot{\varepsilon}_b = 10^{-2}$. The $n_t/n_0$ values are indicated in the figure.

Figure 3 shows the compressive stress for networks with $n_t/n_0 = 10$ at various compression rates. At the slowest rate ($\dot{\varepsilon}_b = 1 \times 10^{-3}$), the stress remains small throughout the compression, indicating that fiber rearrangement via slippage is sufficiently fast against the imposed deformation. As $\dot{\varepsilon}_b$ increases, stress rises progressively. The rate dependence is approximately logarithmic over the range studied. This behavior is consistent with a picture in which the network rearrangement process has a characteristic timescale: when the compression rate exceeds this timescale, the network response transitions from viscous-like flow to a more elastic, jammed state governed by inter-fiber friction and yield-force constraints.

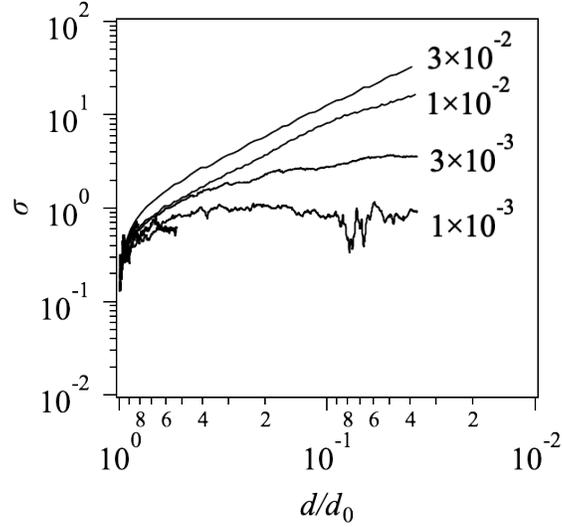

**Figure 3** Stress response under compression for the fiber networks under various compression rates $\dot{\varepsilon}_b$ for the fiber length $n_t/n_0 = 10$. The $\dot{\varepsilon}_b$ values are indicated in the figure.

## 4. Discussion

The classical van Wyk theory [13] and its extensions [14], [15], [16] describe the compression of fiber networks as a quasi-static mechanical-equilibrium problem in which stress arises from elastic bending of fiber segments between contact points. The theory predicts $\sigma \propto E_f \varphi^3$, where $E_f$ is the fiber modulus and $\varphi$ is the volume fraction. Critically, the fiber length does not appear explicitly in the van Wyk framework: at a given volume fraction, the number of contacts per unit volume is determined by the fiber diameter and the volume fraction, regardless of individual fiber length. The proposed simulation, performed at fixed volume fraction, demonstrates a strong dependence of compressive stress on fiber length, with stress varying significantly as seen in Fig. 2. This result arises from the topological nature of fiber networks in the proposed model: longer fibers participate in more inter-fiber contacts than shorter fibers. This physics is fundamentally different from the van Wyk picture of independent beam bending. Furthermore, the van Wyk theory is inherently rate-independent, predicting identical stress responses regardless of the compression speed. In contrast, the simulation results show pronounced rate dependence, with stress varying by more than an order of magnitude across the range of $\dot{\varepsilon}_b$ studied. This rate sensitivity arises from the dissipative nature of fiber slippage: the friction coefficients introduce characteristic timescales, and the yield parameter $F_y$ creates a threshold below which the network can rearrange quasi-statically.

The qualitative trends observed in our simulations are consistent with a substantial body of experimental literature on LFT and GMT compression molding. The rate dependence of compressive stress is a well-established experimental observation. Kotsikos et al. [7] reported that GMT exhibited rate-dependent squeeze-flow behavior. Rienesl et al. [8] also demonstrated shear-thinning behavior in CF-SMC with rate-dependent viscosity. Vahlund and Gebart [26] performed squeeze-flow rheology of GMT at closing velocities up to 30 mm/s and successfully fitted power-law relations between closing velocity and closing force. Schreyer et al. [9] showed that CF-PA6 LFT-D exhibits pronounced shear-rate dependence in squeeze-flow tests, with stress increasing at higher compression velocities. Kelly et al. [27] studied the viscoelastic compression of resin-impregnated fiber networks and reported that stress relaxation could reach up to 64% of the peak stress, with the peak stress increasing at higher strain rates. All of these observations are qualitatively captured by the proposed model, in which the balance between the fiber-rearrangement timescale and the imposed deformation rate governs the stress response.

The fiber-length dependence is less systematically documented, partly because fiber breakage [28] during processing complicates controlled experiments. Nevertheless, Dweib and ÓBrádaigh [10] reported that fiber length affects the rheological properties of E-glass/PP LFTs in squeeze-flow tests. It is also well established in the LFT processing literature that longer fibers require gentler processing conditions (lower shear, compression molding rather than injection molding) to avoid breakage, precisely because longer fibers generate higher flow resistance.

Perhaps the most direct experimental support for the fiber-slippage mechanism comes from the observation of fiber–matrix separation during GMT compression. Huang et al. [12] showed that during squeeze flow, the polymer matrix is pushed outward more readily than the entangled long fibers, causing a progressive increase in fiber volume fraction in the interior of the specimen. They attributed this to fiber contacts and documented a concomitant evolution of the power-law index during compression. This phenomenon is naturally interpreted within the slip-link framework: the yield parameter creates a threshold below which fibers remain locked in place, while the polymer matrix (not modeled here) can flow freely. The separation presumably occurs when the driving stress exceeds the matrix yield stress but remains below the fiber-slippage yield force.

Other relevant experimental systems include non-woven fiber networks and pulp fiber suspensions. Derakhshandeh et al. [29] reviewed the rheology of pulp fiber suspensions, documenting yield stress, rate-dependent viscosity, and viscoelastic behavior that depend strongly on fiber concentration and aspect ratio. The compression mechanics of low-density foam-formed fiber networks have been studied by Ketoja et al. [30], who showed that buckling of fiber segments governs the compression strength, with a square dependence on initial density. While these systems differ in detail from the LFT context, they share the common feature of fiber networks exhibiting rate- and structure-dependent mechanical response.

## 5. Concluding Remarks

This study has demonstrated, for the first time, that the slip-link model that was originally developed and extensively validated for entangled polymer melts can be applied to simulate the compression behavior of long-fiber networks. The model captures two key phenomena that are inaccessible to classical van Wyk-type fiber network compression theories: (i) a strong dependence of compressive stress on fiber length, mediated by the number of contact points per fiber, and (ii) a pronounced dependence on compression rate, arising from the dissipative dynamics of fiber slippage. Both trends are qualitatively consistent with experimental observations from LFT and GMT compression molding.

The principal advantage of the slip-link approach is computational efficiency. By resolving only the contact points rather than the full fiber geometry, the model reduces the degrees of freedom by orders of magnitude compared to direct fiber-level simulations. It should be noted, however, that quantitative comparison with experiments poses a fundamental challenge owing to the non-equilibrium nature of fiber systems. Specifically, the simulation results depend on the initial configuration, which is difficult to reproduce in experimental systems. The determination of relevant parameters, including fiber friction and yield force, also remains challenging, as does the explicit modeling of the matrix. Further tests and developments of the proposed model are ongoing and will be reported elsewhere.

**Author contribution:** The author has accepted responsibility for the entire content of this manuscript and approved its submission.

**Competing interests:** The author states no conflict of interest.

**Research funding:** This study is partly supported by JSPS KAKENHI and the Eno Science Foundation.

**Data availability:** The simulation data that support the findings of this study are available from the corresponding author upon reasonable request.

**Use of AI:** The author used LLM(Claude, Opus4.6) to improve English writing.